\documentclass[12pt]{iopart}
\begin{document}

\title{A discussion on maximum entropy production and information theory}

\author{Stijn Bruers}
\address{Instituut voor Theoretische Fysica, Celestijnenlaan 200D,
 Katholieke Universiteit Leuven, B-3001 Leuven, Belgium}
\ead{stijn.bruers@fys.kuleuven.be}

\begin{abstract}
We will discuss the maximum entropy production (MaxEP) principle based on Jaynes' information theoretical arguments, as was done by Dewar (2003, 2005). With the help of a simple mathematical model of a non-equilibrium system, we will show how to derive minimum and maximum entropy production. Furthermore, the model will help us to clarify some confusing points and to see differences between some MaxEP studies in the literature.
\end{abstract}

\maketitle

\section{Motivation}

The maximum entropy production (MaxEP) is believed to be an organizational principle applicable to physical and biological systems (see reviews in \cite{KleidonLorenz2005, MartiouchevSeleznev2006, OzawaOhmura2003}). There are different attempts to theoretically proof MaxEP. The most detailed mathematical studies were done in two papers by Dewar \cite{Dewar2003, Dewar2005}. Dewar proposed different derivations of MaxEP by using the maximum information entropy (MaxEnt) procedure by Jaynes \cite{Jaynes1957, Jaynes2003}. It is a similar argument as the derivation of the Gibbs ensemble in equilibrium statistical mechanics, but with the crucial difference that the information entropy is not defined by a probability measure on phase space, but on path space. 

In this article, we will comment on the arguments by Dewar. We will do this with a rather simple model which one can easily solve. The most important conclusion is that Dewar discussed basically three different derivations, leading to at least three comments: \\
-The derivation in \cite{Dewar2003} leads in the linear response regime to the well known minimum entropy production (MinEP, \cite{KleinMeijer1954, KondepudiPrigogine1998}) instead of MaxEP.  \\
-The derivation in the main text of \cite{Dewar2005} works only in the linear response regime, and leads to Ziegler's MaxEP principle \cite{Ziegler1983} or a 'linear response' MaxEP principle as used in e.g. \cite{ZupanovicJuretic2004}. \\
-The derivation in the appendix of \cite{Dewar2005} contains some unresolved remarks that need further clarification. We will see that this derivation is related with what we will call the 'total steady state' MaxEP. \\

Furthermore, Dewar refers to the work by e.g. Paltridge \cite{Paltridge1979} on climate systems. We will demonstrate that Paltridge's MaxEP principle, which we will call 'partial steady state' MaxEP, is different from the above mentioned MaxEP principles in. Hence. 'partial steady state' MaxEP is unrelated with the principles derived in \cite{Dewar2003, Dewar2005}. This will lead us to the important conclusion that there are different MaxEP principles discussed in the literature, often leading to some confusion. In fact, some studies \cite{OzawaShimokawa2001, ShimokawaOzawa2002}, in particular in fluid dynamics, discussed even another principle, what we will call the 'non-variational MaxEP' principle. In the appendix we point at a rough analogy between Paltridge's 'partial steady state' MaxEP principle and an equilibrium system with MaxEnt. There are some theoretical problems associated with this analogy, but nevertheless we present it to clarify the line of reasoning of MaxEP and its possible information theoretical derivation.

\section{The model}

Let us consider a system of $\iota$ sites, with a real variable $n_i(t)$ ($i=1,\,..,\iota)$ at each site. These variables depend on the discrete time $t=0,\,1,..,\tau$. At every timestep there is a random flux between the sites. The flux $f_{ij}=-f_{ji}$ from $i$ to $j$ depends on a real constant parameter $c_{ij}=c_{ji}$, such that $f_{ij}(t)=\pm c_{ij}$ where the sign is stochastic. A microscopic path $\Gamma$ is a specific set of values $+c_{ij}$ or $-c_{ij}$ for every timestep and every $i$ and $j$. The pathspace is the set of all possible paths. The sign stochasticity gives a stochastic dynamics, such that for each microscopic path we have 
\begin{eqnarray}
n_{i,\Gamma}(t+1)-n_{i,\Gamma}(t)=-\sum_j f_{ij,\Gamma}(t). \label{eom}
\end{eqnarray}

The time averages depend on the path $\Gamma$ and are denoted with an overline, e.g. $\overline{f_{ij,\Gamma}}=\frac{1}{\tau}\sum_t f_{ij}(t)$. For each microscopic path, we assign a probability $p_{\Gamma}$. The path ensemble averages are written with brackets, e.g. $\left\langle \overline{f_{ij}}\right\rangle \equiv \sum_\Gamma p_\Gamma \overline{f_{ij,\Gamma}}$. 
 
To find the most likely probability measure on path space, we will use Jaynes' information theory formalism by maximizing the path information entropy 
\begin{eqnarray}
S_I\equiv -\sum_\Gamma p_\Gamma \ln p_\Gamma \label{S_I}
\end{eqnarray}
under the constraints 
\begin{eqnarray}
&&\sum_\Gamma p_\Gamma =1,\label{constraint 1}\\
&&\sum_\Gamma p_\Gamma n_{i,\Gamma}(0)=\left\langle n_i(0) \right\rangle, \label{constraint n}\\
&&\sum_\Gamma p_\Gamma \overline{f_{ij,\Gamma}}=F_{ij}, \label{constraint F}
\end{eqnarray}
for some (or all) $i$ and $j$. These constraints were used by Dewar \cite{Dewar2003} and are to be interpretted as follows:\\
The first constraint is the normalization of the probability measure.\\
The second constraint means that at the initial time, the (path ensemble average) value of $n_i$ is measured. $n_{i,\Gamma}(0)$ is not dependent on the complete path, but only on the initial time value of the path.\\
The third constraint means that the time and path ensemble average of the flux from $i$ to $j$ is measured to be the numerical value $F_{ij}$.

The maximum of $S_I$ under the constraints results in 
\begin{eqnarray}
p_\Gamma=\frac{1}{Z}\exp{A_\Gamma} \label{p_Gamma}
\end{eqnarray}
with the path action
\begin{eqnarray}
A_\Gamma = \sum_i \lambda_i n_{i\Gamma}(0)+\sum_{ij}\eta_{ij}\overline{f_{ij,\Gamma}},
\end{eqnarray}
with $\lambda_i$ and $\eta_{ij}=-\eta_{ji}$ Lagrange multipliers of constraints (\ref{constraint n}) and (\ref{constraint F}) respectively. By deriving 
\begin{eqnarray}
\left\langle n_{i}(0)\right\rangle=\frac{\partial \ln Z}{\partial \lambda_i}, 
\end{eqnarray}
and using (\ref{constraint n}), we get $\left\langle n_{i}(0)\right\rangle=n_{i,\Gamma}(0)$. This basically means that constraint (\ref{constraint n}) is trivially satisfied due to (\ref{constraint 1}), so we can take $\lambda_i=0$. The reason behind this is that $n_i(0)$ did not depend on the complete path.

The partition sum and the Lagrange multipliers $\eta_{ij}$ can be easily calculated:
\begin{eqnarray}
&&Z=\prod_{ij}Z_{ij}=\prod_{ij}(2\cosh \frac{\eta_{ij}c_{ij}}{\tau})^\tau, \label{Z}\\
&&\frac{\partial \ln Z}{\partial \eta_{ij}}=F_{ij}=c_{ij}\tanh(\frac{\eta_{ij}c_{ij}}{\tau}), \label{F}\\
&&X_{ij}\equiv \frac{\eta_{ij}}{\tau}=\frac{1}{c_{ij}}\mathrm{arcth} \frac{F_{ij}}{c_{ij}}. \label{X}
\end{eqnarray}
One can split the time and ensemble averaged fluxes $F$ in forward and a backward components: 
\begin{eqnarray}
F_{ij}&=&F_{ij}^+-F_{ij}^-\\
&=&\frac{c_{ij}e^{X_{ij}c_{ij}}}{e^{X_{ij}c_{ij}}+e^{-X_{ij}c_{ij}}}-\frac{c_{ij}e^{-X_{ij}c_{ij}}}{e^{X_{ij}c_{ij}}+e^{-X_{ij}c_{ij}}},
\end{eqnarray}
such that $2c_{ij}X_{ij}=\ln(\frac{F_{ij}^+}{F_{ij}^-})$ (which is a well known expression for the thermodynamic forces for e.g. elementary chemical reactions \cite{KondepudiPrigogine1998}).
(\ref{X}) are the constitutive (phenomenological) equations of motion.

Notice that our description of a stochastic non-equilibrium dynamical model is mathematically equivalent with a statistical equilibrium ferromagnetic spin model. This can be seen by interpreting the fluxes $f_{ij}(t)$ as the values of the ferromagnetic spins $s_{ij,t}$. These spins take values $\pm c_{ij}$ for every $t$. Instead of time, $t$ is interpretted as a spatial coordinate, so for every $i$ and $j$ we have a one dimensional spin chain. The observed averaged fluxes $F_{ij}$ correspond with the observed mean magnetisations $m_{ij}$ for every spin chain. 
In the equilibrium spin model, the multipliers $\eta$ are basically inverse temperatures of the chains. In the non-equilibrium interpretation, these multipliers are related with the thermodynamic driving forces $X$ which are conjugate to the fluxes $F$.

As in \cite{Dewar2003}, the entropy production (EP) of a microscopic path $\Gamma$ will be defined as the time antisymmetric (irreversible) part of the action, written as $\sigma_\Gamma\equiv A_\Gamma^{irr}/\tau$. In our example, the fluxes are all time antisymmetric and there is no symmetric (reversible) part of the action, so we have $\sigma_\Gamma=\sum_{ij}X_{ij}\overline{f_{ij,\Gamma}}$. The expectation value of the EP is (for convenience written without brackets)
\begin{eqnarray}
\sigma\equiv\sum_\Gamma p_\Gamma \sigma_\Gamma = \sum_{ij}X_{ij}\sum_\Gamma p_\Gamma \overline{f_{ij,\Gamma}}=\sum_{ij}X_{ij}F_{ij},
\end{eqnarray}
which is the classical expression for the EP as a bilinear form of forces and fluxes.

Plugging the solution (\ref{p_Gamma}) into (\ref{S_I}), we get the maximum information entropy as a function of the forces: 
\begin{eqnarray}
S_{I,max}(X)&=&\ln Z(X)-\left\langle A(X)\right\rangle \\
&=&\ln\left( \prod_{ij}\frac{2\cosh(X_{ij}c_{ij})}{\exp(X_{ij}c_{ij}\tanh(X_{ij}c_{ij}))}\right) ^\tau,\\
&\approx& \ln W(\left\langle A(X)\right\rangle ) \label{W}
\end{eqnarray}
with $W(\left\langle A(X)\right\rangle)$ the 'density of paths': The number of paths with approximately the average $\left\langle A(X)\right\rangle$ as path action.

Next we introduce the entropy curvature (or response) matrix as in \cite{Dewar2005}
\begin{eqnarray}
A_{ij,kl}(F)&\equiv& \frac{\partial X_{ij}}{\partial F_{kl}} \label{A_ij,kl}\\
&=&-\frac{\partial^2 S_{I,max}(X(F))}{\tau\partial F_{ij}\partial F_{kl}}\\
&=&\delta_{ij,kl}\frac{1}{c_{kl}^2-F_{kl}^2},
\end{eqnarray}
with $\delta_{ij,kl}$ the Kronecker delta matrix ($\delta_{ij,kl}=1$ iff $i=k$ and $j=l$). 

With the steepest descent approximation (i.e. a quadratic expansion around average $F$), as in \cite{Dewar2005}, we can calculate the probability distribution for the time averaged flux
\begin{eqnarray}
p(\overline{f})\propto \exp \left( -\frac{\tau}{2}\sum_{ij,kl}[\overline{f_{ij}}-F_{ij}]A_{ij,kl}(F)[\overline{f_{kl}}-F_{kl}]\right) . \label{p(f)}
\end{eqnarray}
Combining this expression with the the fluctuation theorem \cite{EvansCohen1993, EvansSearles2002}
\begin{eqnarray}
\frac{p(\overline{f})}{p(-\overline{f})}=\exp(2\tau \sigma(\overline{f}))=\exp(2\tau\sum_{ij}X_{ij}\overline{f_{ij}}), \label{fluctuationtheorem}
\end{eqnarray}
and taking together the terms linear in $\overline{f}$ in the exponent, Dewar \cite{Dewar2005} derived another expression for the constitutive equation (compare with (\ref{X})):
\begin{eqnarray}
X_{ij}=\sum_{kl}A_{ij,kl}(F)F_{kl}. \label{X=AF}
\end{eqnarray}
Below, we will point at some hidden assumption in this derivation, clarifying the difference between (\ref{X}) and (\ref{X=AF}).

As a final definition, we introduce the dissipation function as in \cite{Dewar2005}
\begin{eqnarray}
D(F)\equiv 2\sum_{ij,kl}A_{ij,kl}(F)F_{ij}F_{kl}.
\end{eqnarray}

In the linear response regime near thermodynamic equilibrium, all forces $X$ are small and by (\ref{X}) they are (approximately) linearly related with the fluxes as
\begin{eqnarray}
X_{ij,lin}\approx F_{ij}/c_{ij}^2 \label{Xlin}
\end{eqnarray}
In this regime, the two constitutive equations (\ref{X}) and (\ref{X=AF}) become equal to (\ref{Xlin}), and the dissipation function (approximately) equals the EP 
\begin{eqnarray}
D_{lin}(F)\approx \sigma(X(F),F).\label{Dlin}
\end{eqnarray} 

This is the basic set-up, as discussed in \cite{Dewar2003, Dewar2005}. Now we will give some comments on Dewar's arguments. 

\section{Comments}

\subsection{Linear response minimum entropy production} \label{subsec MinEP}
The first article \cite{Dewar2003} focused on the non-equilibrium steady state. Up till now, the forces $X_{ij}$ (and the parameter values $c_{ij}$) were supposed to be constants. However, in most systems, they can change. Let us introduce a new, longer timescale $T\equiv t/\tau$. The forces, fluxes and parameters are approximately constant for short timescales $0\leq t \leq \tau$, but they can slowly change. Suppose the system $X(T),\, F(T)$ attains a steady state for $T\rightarrow \infty$. What happens with the EP $\sigma(T)$?

As can be seen by the counting argument (\ref{W}), we can calculate $W$ in the linear response regime for small $X$:
\begin{eqnarray}
W(\left\langle A(X)\right\rangle)\approx \left[ \prod_{ij}(2-(X_{ij}c_{ij})^2)\right] ^\tau\approx \left[ 2^{\iota^2-\iota}-2^{\iota^2-\iota-1}\sum_{ij}(X_{ij}c_{ij})^2\right] ^\tau
\end{eqnarray}
with this simplification, the path information becomes
\begin{eqnarray}
S_{I,max}(X)&=&\ln(W)\approx \ln 2^{\tau(\iota^2-\iota)}+\tau\ln\left( 1-\frac{\sigma}{2}\right)\\
&\approx& \ln 2^{\tau(\iota^2-\iota)}-\frac{\tau \sigma}{2}.
\end{eqnarray}
The interpretation of this result is clear: The first term on the right hand side is the logarithm of the total number of paths (for a uniform probability distribution). The second term only contains the EP. In \cite{Dewar2003}, an important assumption was made in order to derive MaxEP: The number of paths $W$ should be an increasing function of the averaged action. This averaged action is proportional with the EP, and hence one could claim that the higher the EP, the higher $S_{I,max}$. However, here we obtain the reverse, resulting in a minimization of the entropy production (MinEP). Suppose there are additional constraints such as 
\begin{eqnarray}
\sum_{ij}\beta_{ij,e}X_{ij}(T)=X_e^0 \label{forcesconstraintgeneral}
\end{eqnarray}
with $X_e^0$ constant 'external' forces. An example of this kind of constraint on the forces is the Kirchoff loop law in electrical networks. Minimizing the EP $\sigma$ under these constraints, and using the linearized constitutive equations (\ref{Xlin}), one can find (for $T\rightarrow\infty$) the unique steady state (written with $*$), which for site $i$ is given by 
\begin{eqnarray}
\sum_j F_{ij}^*=0 \label{steadystateconstraintgeneral}
\end{eqnarray}
(as follows from (\ref{eom})). We conclude that the derivation in \cite{Dewar2003} can be used to derive MinEP rather than MaxEP, because the assumption that $W$ is a decreasing function of $\sigma$ is valid in our model.


\subsection{Ziegler's and linear response MaxEP} 
Let us now comment on \cite{Dewar2005}. Ziegler \cite{Ziegler1983} has proposed a MaxEP principle to derive the constitutive equations. It only works for systems in the linear response regime (and some highly restricted exceptional cases mentioned in \cite{Ziegler1983}, but we will not discuss them here). It is variational principle, with Lagrangian which is to be maximized:
\begin{eqnarray}
\mathcal L_{Ziegler} (F)\equiv D(F)+\gamma(D(F)-2\sum X_{ij}^0 F_{ij}). \label{L_Ziegler}
\end{eqnarray}
The last term is a constraint with Lagrange multiplier $\gamma$. In this variation, $X^0$ is kept fixed. It is this 'maximum dissipation' principle that was explained in \cite{Dewar2005}, eq. (22)\footnote{In \cite{Dewar2005}, a 'dual' version is applied, switching the roles of the forces and the fluxes. It is mathematically equivalent with our formulation.}. 

It is important to keep in mind that (contrary to what is claimed in \cite{Dewar2005}) the constitutive equations derived from the above Lagrangian are only compatible with (\ref{A_ij,kl}) and (\ref{X=AF}) when $\sum_{ij} F_{ij}\frac{\partial A_{ij,kl}}{\partial F_{lm}}=0$. It is clear that this restriction does not hold in the non-linear regime of our model with constitutive equations (\ref{X}). Only in the linear response regime (when $A$ is a constant matrix, leading to (\ref{Xlin})) is (\ref{X=AF}) compatible with (\ref{X}). The reason why the derivation in \cite{Dewar2005} only works near equilibrium (i.e. in the linear response regime) is due to the use of a steepest descent approximation in (\ref{p(f)}). This works only when the fluxes $f_{ij}$ are close to their expectations $F_{ij}$. But using the fluctuation theorem (\ref{fluctuationtheorem}), also $-f_{ij}$ should be close to $F_{ij}$. This is only possible when $F_{ij}$ is small.

The derivation in \cite{Dewar2005} has also another application, as one can add more constraints to the above Ziegler's principle in order not to find the constitutive equations, but to find the unique near-equilibrium steady state. This is also a MaxEP principle, which we will name 'linear response MaxEP' because it only works in the near-equilibrium linear response regime.\footnote{As was correctly noted in \cite{Dewar2005}, this principle should not to be confused with the linear response minimum entropy production principle \cite{KondepudiPrigogine1998}, which uses other constraints resulting in a minimum of the EP at the near-equilibrium steady state.} Zupanovi\'c et al. \cite{ZupanovicJuretic2004} discussed this principle with an electrical network as an example, whereby the forces are the voltages. The Lagrangian generally looks like
\begin{eqnarray}
\mathcal L_{linear} = D(F)+\gamma_0 (D(F)-\sum_e X_e^0 F_e)+\sum_e\gamma_e(\sum_{ij}\beta_{ij,e}F_{ij}-F_e). \label{L_linear}
\end{eqnarray}
The second term on the right hand side is the constraint which says that in the steady state the power influx into the system due to the fixed external driving forces $X_e^0$ (with conjugate external fluxes $F_e$ that do not contribute to the dissipation $D(F)$) is completely dissipated. In \cite{ZupanovicJuretic2004}, this fixed external driving force is the applied voltage of a battery, and $F_e$ is the current through this battery. The last term (with constants $\beta_{ij,e}$) is a steady state constraint on the fluxes. In the electrical network example in \cite{ZupanovicJuretic2004} it is the Kirchoff's current law.

Note that, as in the previous comment, we can use a counting argument to derive Ziegler's or linear response MaxEP. In the near equilibrium regime we have (\ref{Dlin}) and $W(\sigma)=W(D)$ becomes maximal under the constraints in (\ref{L_Ziegler}) or (\ref{L_linear}).

\subsection{Partial steady state MaxEP} \label{subsecPaltridge}
In his two papers, Dewar also refers to the work by Paltridge that gives experimental validation of the MaxEP principle. The basic idea of the climate model of Paltridge is similar to the idea in e.g. \cite{DewarJuretic2006} or \cite{JureticZupanovic2003} for chemical reactions. Paltridge divides the universe in compartments (sun, equator, pole and deep space) with energy fluxes between them, just as the chemical reaction system of ATP synthase in \cite{DewarJuretic2006} consists of compartments (the different molecular states) with particle fluxes between them. In the Paltridge model, there is atmospheric heat transport from equator to pole, and its transport coefficient is a priori not known. This coefficient is guessed by maximizing the EP associated with the atmospheric heat transport processes. The other processes and parameters related with the heat radiation (e.g. from sun to equator) are a priori known, and the earth system is supposed to be in the steady state. Note that not the total EP is maximized. In \cite{DewarJuretic2006}, a parameter $\kappa$ and the flux $F(\kappa)$ between the compartments O:ATP and O:P.ADP are unknown. The most likely values for this parameter and flux are derived by maximizing the corresponding EP (not the total EP of all reactions), knowing that the system is in the steady state.

Making the analogy with our model, we can take a system consisting of three compartments (sites with $\iota=3$), with parameters $c_{13}=0$ and $c_{12}\neq 0$ a priori known. As the atmospheric heat transport coefficient or the $\kappa$ parameter, $c_{23}$ is unknown, and it is guessed by maximizing the corresponding partial EP $\sigma_{23}=X_{23}F_{23}$ under the steady state conditions. This explains the name 'partial steady state' MaxEP. The steady state conditions are e.g. 
\begin{eqnarray}
X_{12}+X_{23}=X_e^0 \label{totalforceconstraint}
\end{eqnarray}
(as a specific example of (\ref{forcesconstraintgeneral})) with $X_e^0$ known and fixed, and 
\begin{eqnarray}
F_{12}^*=F_{23}^*. \label{steadystateconstraint}
\end{eqnarray}
(This is the steady state condition for the middle site $2$, as a specific example of (\ref{steadystateconstraintgeneral}). The total system, including sites $1$ and $3$, is not in the steady state, except when the total system is in equilibrium.) 

Under these constraints, the partial EP can be written as
\begin{eqnarray}
\sigma_{23}^*=F_{23}^*\left( X_e^0-\frac{1}{c_{12}}\mathrm{arcth}\frac{F_{23}^*}{c_{12}}\right) . \label{EP23}
\end{eqnarray}
The maximum gives a complicated expression of $F_{23,max}^*(X_e^0,c_{12})$ as a function of the known parameters. This also gives $c_{23,max}(X_e^0,c_{12})$. Although it is believed \cite{JureticZupanovic2003, Paltridge1979, OzawaOhmura2003} that this principle is applicable to the far-from-equilibrium regime, we can also look at the linear response regime, where it is easy to calculate that $c_{23,max}=c_{12}$ and $F_{23,max}^*=\frac{c_{12}^2}{2}X_e^0$.

As mentioned above, \cite{Dewar2003} results in minimum EP and \cite{Dewar2005} results in Ziegler's or linear MaxEP, and these principles are different in nature than Paltridge's MaxEP principle discussed here. In the appendix we give an analogy of our model with an equilibrium model. Although theoretically not very rigid, the discussed analogy might serve as a general guideline to clarify the partial steady state MaxEP. For the moment, it is important to stress that this MaxEP principle remains an unproven hypothesis with a lot of controversy and unsolved questions about the necessary conditions, requirements and ranges of application.

\subsection{Total steady state MaxEP} \label{subsecappendix}
In the appendix of \cite{Dewar2005}, Dewar gives a third information theoretical derivation for MaxEP. An important assumption is made: The total dissipation or (more generally) the total entropy production should have an upper bound $\sigma(X(F),F)\leq \sigma_{max}$ under some prior information $C$ (such as the knowledge that the system is in the steady state (\ref{totalforceconstraint}-\ref{steadystateconstraint})). Looking at the example in section \ref{subsecPaltridge}, in the steady state in the linear response regime, the EP becomes
\begin{eqnarray}
\sigma^*=X_e^0 F_{12}^*=\frac{c_{12}^2c_{23}^2}{c_{12}^2+c_{23}^2}(X_e^0)^2 \leq c_{12}^2(X_e^0)^{2}.
\end{eqnarray}
The maximum is attained for $c_{23}\rightarrow \infty$. This is not compatible with $c_{23,max}=c_{12}$ obtained by maximizing the entropy production of the unknown flux in section \ref{subsecPaltridge}. As we varied the total EP in the steady state with respect to an unknown parameter, this explains the chosen name 'total steady state MaxEP'.

One can place questions about the choice of constraints used in Dewar's appendix derivation. Why not add the inequality constraint $\sigma\geq 0$ as a consequence of (\ref{fluctuationtheorem}), or the steady state constraints (\ref{totalforceconstraint}-\ref{steadystateconstraint})? And is the obtained probability measure a maximum of the information entropy? We will not deal with these questions here, as they should be taken up in future work.

\subsection{Non-variational MaxEP}

At the end of his paper, Dewar \cite{Dewar2005} mentions the Rayleigh-B\'enard convective fluid system. Others (e.g. \cite{OzawaShimokawa2001, ShimokawaOzawa2002}) have made a MaxEP hypothesis for other fluid systems. We will call this principle the 'non-variational' MaxEP, because contrary to the above mentioned principles, it is a selection principle rather than a variational principle varying the EP with respect to some continuous variable. 

Suppose a system has a highly non-linear dynamics, resulting into the possibility of having a discrete set of steady states. The hypothesis claims that the selected state (e.g. the most stable) is the one with highest EP of all the steady states. E.g. in the Rayleigh-B\'enard system, the steady states are a heat conduction state, a heat convection state and perhaps other (turbulent) states. For temperature gradient values beyond a critical transition point, the heat convection state is most stable, and it has the highest heat transport and the highest EP (see also \cite{SchneiderKay1994b}).

Making the analogy with our model, we will take a time dependence $c_{ij}(T)$ as in section \ref{subsec MinEP}. This might give a non-linear dynamics, resulting into different steady states for the fluxes $F_{ij}^*$. The hypothesis will be proven when the most (asymptotically) stable state has the highest EP. Up till now, no proof of this hypothesis is known, and it is doubtful whether it is generally true. 

\subsection{Microscopic MaxEP}
As a smaller final comment, our model demonstrates another kind of MaxEP principle, different from the above principles. One might look for the microscopic path which has the highest probability (\ref{p_Gamma}). In our model, we can easily see that this path should have the maximum value of the action $A_{\Gamma,max}=\tau\sum_{ij} c_{ij}|X_{ij}|$. This corresponds with a maximum of the microscopic path EP $\sigma_\Gamma$. This microscopic path EP does not necessarily result in a maximum of the path-ensemble averaged EP $\sigma$. Furthermore, there are some doubts that this 'microscopic MaxEP' is generally true (Maes, private communication). We have seen that the action $A_\Gamma$ in our model is basically the time-antisymmetric part, which is the EP \cite{Maes2003}. But in more general descriptions for other systems, there is also a time-symmetric part of the action \cite{Maes2003}. When this part also depends on the path, the microscopic MaxEP might be unvalid.

\appendix

\section*{Appendix: A MaxEP-MaxEnt analogy}

In section \ref{subsecPaltridge}, we have described the partial steady state MaxEP principle with a simple example. The intuition of Dewar and others is that this MaxEP principle can be derived by maximizing the path information entropy in non-equilibrium statistical mechanics, the same way that Jaynes \cite{Jaynes1957, Jaynes2003} derived the Gibbs probability measure in equilibrium statistical mechanics, by maximizing the phase space information entropy. This method is called MaxEnt. 

Here we will discuss an analogy of this non-equilibrium MaxEP system with an equilibrium MaxEnt system, in order to clarify the line of reasoning used in this MaxEP principle. The analogy below is very rough, and definitely not a proof for MaxEP. There are a lot of theoretical problems with it, so one should not take it to serious.

The non-equilibrium MaxEP: Take a system consisting of three compartments with two fluxes between them.
Let us take the linear response regime, where these fluxes $F_{ij}$ have conductances $C_{ij}=c_{ij}^2$ relating the forces $X_{ij}=F_{ij}/C_{ij}$. Suppose the conductance $C_{23}$ is unknown. This means that also the steady state values (using (\ref{totalforceconstraint}-\ref{steadystateconstraint})) of $X_{ij}$ and $F_{ij}$ are unknown. MaxEP claims that they can be derived by maximizing the partial steady state EP (\ref{EP23}) $\sigma_{23}^*=F_{23}^*(X_e^0-F_{23}^*/C_{12})$. 

The equilibrium MaxEnt: Consider a closed system (energetically coupled with an environment), consisting of two closed boxes which are also energetically coupled. For simplicity, the volumes and heat capacities of the two boxes are equal to unity. The two boxes contain an ideal gas with particle numbers $N_1$ and $N_2$ at temperatures $T_1$ and $T_2$ respectively. Suppose that a priori only $T_1$ and the total number of particles $N^0=N_1+N_2$ are known and constant. The other variables and parameters are derived by MaxEnt.

The following table represents the analogy schematically:

\begin{tabular}{|c|c|}
  \hline
  MaxEP & MaxEnt \\
  \hline
  \hline
  fluxes: $F_{12},\, F_{23}$ & energies: $E_1,\, E_2$\\
  \hline
  conductances: $C_{12},\, C_{23}$ & temperatures: $T_1,\, T_2$\\
  \hline
  forces: $X_{12},\,X_{23}$ & particle numbers: $N_1,\,N_2$\\
  \hline
  linear response approximation: & ideal gas approximation: \\
  $X_{ij}=F_{ij}/C_{ij}$ & $N_i=E_i/T_i$\\
  \hline
  steady state: $F_{12}^*=F_{23}^*$ & energy equality: $E_1^*=E_2^*$\\
  \hline
  non-equilibrium constraint: & particle conservation:\\
  $X_{12}+X_{23}=X_e^0$ is constant & $N_1+N_2=N^0$ is constant\\
  \hline
  unknown: $F_{ij}, C_{23}, X_{ij}$ & unknown: $N_i, T_2, E_i$\\
  \hline
  MaxEP$\rightarrow$ $C_{23,Max}=C_{12}$ & MaxEnt$\rightarrow$ $T_{2,Max}=T_1$\\
  $F_{ij,Max}^*=\frac{X_e^0 C_{12}}{2}$ & $E_{i,Max}^*=\frac{N^0 T_1}{2}$ \\
  \hline
\end{tabular}
\\
\\
Off course, one can always take a system with different conductances, so MaxEP is not generally true. A similar possibility occurs in the well known equilibrium statistical physics: When the two boxes in the MaxEnt system are energetically isolated, it is also not necessary that $T_2=T_1$. As energetic coupling is a necessary condition for temperature equilibration in the MaxEnt formulation, there should be an analogous necessary condition in the non-equilibrium system in order that MaxEP is valid. Once one can find this kind of 'coupling' in the non-equilibrium system, and once one can demonstrate that the path information entropy is (perhaps under some further restrictions) related with the partial EP corresponding with an unknown parameter, then one can give a best guess for this parameter. In this way, perhaps the best guess for e.g. the atmospheric heat conduction parameter in the Paltridge model is derived by maximizing the atmospheric EP. 

The above discussion might give a hint to explain why the experimental atmospheric heat transport is close to the MaxEP value. Dewar \cite{Dewar2005} correctly pointed out that the predictive success of MaxEnt hinges on having correctly identified the constraints. As the temperature equality in the two box system depends on the energetic coupling due to the absence of internal constraints (e.g. dividing isolating walls), the MaxEP heat transport value might perhaps also depend some coupling due to the absence of constraints (e.g. the conductances should be sufficiently variable). We end this appendix by repeating that the above ideas are still very speculative.

\section*{Acknowledgments}
The author wishes to thank R. Dewar, C. Maes and an anonymous referee for helpful comments.

\end{document}